\documentclass[12pt,preprint]{aastex}

\newcommand{\exhydra}{EX~Hydrae}
\newcommand{\exhya}{EX~Hya}
\newcommand{\phispin}{\ifmmode{\phi_{67}}\else{$\phi_{67}$}\fi}
\newcommand{\phiorbit}{\ifmmode{\phi_{98}}\else{$\phi_{98}$}\fi}
\newcommand{\phispinfold}{\ifmmode{\phi_{67,{\rm fold}}}\else{$\phi_{67,{\rm fold}}$}\fi}
\newcommand{\phiorbitfold}{\ifmmode{\phi_{98,{\rm fold}}}\else{$\phi_{98,{\rm fold}}$}\fi}
\newcommand{\kms}{\ifmmode{{\rm km~s}^{-1}}\else{km~s$^{-1}$}\fi}

\shorttitle{Velocity and Mass of EX Hya Measured with Chandra}
\shortauthors{Hoogerwerf et al.}

\begin{document}


\title{The Radial Velocity and Mass of the White Dwarf of EX Hydrae
Measured with Chandra}


\author{R.\ Hoogerwerf, N.~S.\ Brickhouse} 
\affil{Smithsonian Astrophysical Observatory, Harvard-Smithsonian
Center for Astrophysics, 60 Garden Street, MS 31, Cambridge, MA 02138}
\email{rhoogerwerf@cfa.harvard.edu,nbrickhouse@cfa.harvard.edu}

\and

\author{C.~W.\ Mauche}
\affil{Lawrence Livermore National Laboratory, L-473, 7000 East Avenue,
Livermore, CA 94550}
\email{mauche@cygnus.llnl.gov}


\begin{abstract}
We present the first detection of orbital motion in the cataclysmic
variable \exhydra\ based on X-ray data from the {\it Chandra X-ray
Observatory}. The large collecting area of the telescope and the high
resolution of the HETG spectrometers allow for an unprecedented
velocity accuracy of $\sim$15 \kms\ in the X-ray wavelength regime. We
find an emission line velocity amplitude of 58.2$\pm$3.7~\kms \ and
infer a white dwarf mass of $0.49\pm0.13~{\rm M}_\odot$, in good
agreement with previous studies using optical, ultraviolet, and far
ultraviolet data.
\end{abstract}




\section{Introduction}
Magnetic cataclysmic variables (CVs) are semi-detached, interacting
binaries consisting of a late-type star that
overflows its Roche-lobe
and a white dwarf with a magnetic field sufficiently strong to
influence the accretion of matter originating from the late-type
companion.  The magnetic field directs the accretion flow away
from the orbital plane and onto both of the white dwarf's magnetic
poles. Accretion behaves differently in the two classes of magnetic
cataclysmic variables. In polars the stars are tidally locked; this
``static'' orientation results in an accreting flow from the
Lagrangian point L1 of the binary along the magnetic field lines of
the white dwarf onto its magnetic pole(s). In intermediate polars (IPs),
where the white dwarf is spinning faster than the orbital period,
matter also accretes onto the white dwarf along the magnetic field
lines; however, the matter first accumulates in an accretion disk
surrounding the white dwarf. In both polars and IPs the accreting
matter approaches the poles of the white dwarf supersonically and so
passes through a stand-off shock before cooling and settling onto
the white dwarf \citep{aiz1973,kyl1982}.

The cataclysmic variable \exhydra\ belongs to the IP class. It
consists of a white dwarf with a spin period of 67 minutes and has an
orbital period of 98 minutes. Both the spin period and binary period
can be identified in its light curves, from optical
\citep*[e.g.,][]{vog1980,ste1983}, UV
\citep*[e.g.,][]{mau1999,bel2003}, and X-ray emission
\citep*[e.g.,][]{wat1978,all1998}. Furthermore, the inclination of the
\exhya\ system is such that its light curve shows a partial eclipse of
the lower accretion pole of the white dwarf by the companion
\citep{all1998}.

Optical and UV spectroscopy of the \exhya\ system has detected the
orbital motion associated with the white dwarf
\citep*[e.g.,][]{bre1980,hel1987,mau1999,bel2003}. This motion,
combined with the orbital velocity of the companion and the presence
of the eclipse, allows for a mass determination of the white dwarf
independent of models of the accretion column.  In this letter we use
{\it Chandra} High-Energy Transmission Grating (HETG) data for \exhya\
and show that its orbital motion can be detected with high resolution
X-ray spectra.

\section{Observation and Reduction}

\exhya\ was observed by {\it Chandra} using the HETG in combination
with the Advanced CCD Imaging Spectrometer in its spectroscopy layout
(ACIS-S) on 2000 May 18 for 60 ks.  The observation was continuous and
covers $\sim$10 orbital revolutions of the binary system and $\sim$15
white dwarf revolutions.

We reduced the data using the {\it Chandra} Interactive Analysis of
Observations (CIAO) software
package\footnote{http://cxc.harvard.edu/ciao/} making only two
departures from the standard reduction: 1) we turned randomization off
(i.e., we set {\tt rand\_pix\_size} = 0.0 in the {\tt
tg\_resolve\_events} tool) to minimize any artificial line broadening,
and 2) we applied a solar system barycentric correction so that the
event times are in Barycentric Dynamical Time instead of spacecraft
time. The reduced HETG \exhya\ spectrum, as shown in Figure~1 (see
also \citealt{mau2002,muk2003}), reveals emission lines from H- and
He-like ions, as well as from Fe L-shell ions, superimposed on a weak
continuum. \ion{Fe}{17} and \ion{Fe}{22} density diagnostics reported
by \citet*{mau2001,mau2003} indicate densities of order $n_e \approx
10^{14}$ cm$^{-3}$.

\section{Velocity signature}
To search for velocity signatures in the \exhya\ HETG data we used the
following procedure: 1) identify the spectral lines in the first-order
High Energy Grating (HEG) and Medium Energy Grating (MEG) spectra; 2)
divide the observation into time segments; 3) add the spectral lines
in each time segment to form a composite line profile (CLP) for each
grating arm in order to boost the signal-to-noise ratio; 4) measure
the line centroid of the CLP; and 5) look for variations of the line
centroid with orbital or white dwarf spin phase.

\subsection{Line Identification and Wavelength Calibration}
Table~\ref{table_lines} lists the isolated spectral lines, i.e., lines
that do not have another strong line within two times the instrumental
FWHM, with a signal-to-noise ratio larger than 10. We obtained the
reference wavelengths for these lines from the Astrophysical Plasma
Emission Database \citep[APED;][]{smi2001}. Reference wavelengths with
uncertainty estimates from APED are taken from quantum electrodynamic
theory for H-like ions \citep{eri1977} and He-like ions
\citep{dra1988} and from laboratory measurements for Fe L-shell lines
\citep{bro1998, bro2002}. Wavelength uncertainties are for individual
lines only and do not include the effects of dielectronic satellite
lines that are associated with many of these strong lines. Such
blending satellite lines contribute less than a few percent to the
strength of the emission feature. 

We defined unblended lines as either single lines or multiplets that
are unresolved and from the same ion. For unresolved lines from the
same ion we used the emissivity-weighted wavelengths. The He-like Mg
and Si forbidden and intercombination lines are unresolved by MEG, and
since their theoretical line flux ratios depend on electron density
and photoexciting radiation, we cannot weight their wavelengths by the
low-density emissivities tabulated in APED. Thus we excluded theses
lines from the CLP.

\begin{figure}
\plotone{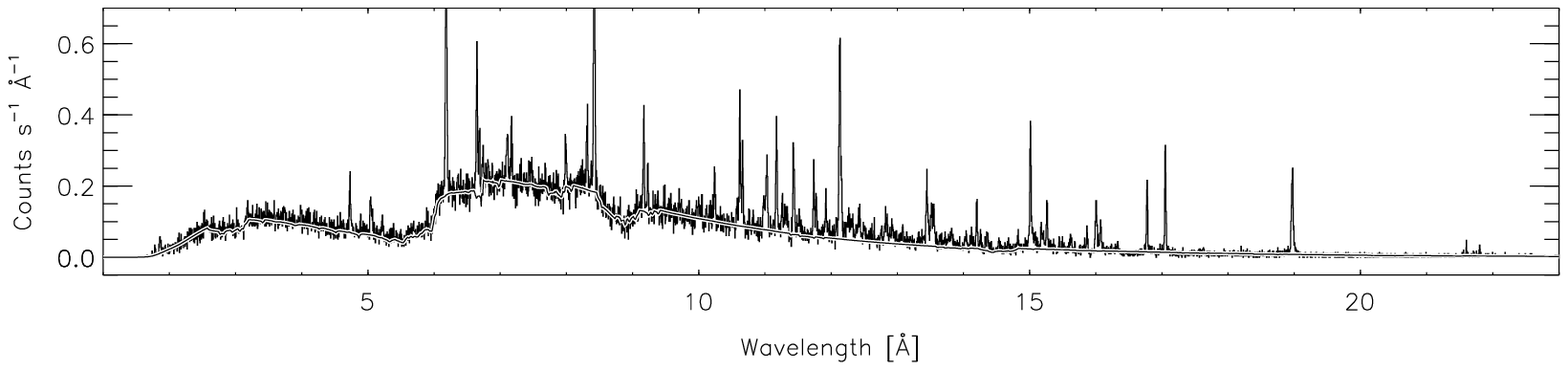}
\caption{Spectrum for the combined first orders of the MEG instrument.
Smooth curve represents the one-temperature thermal
continuum model fit to the spectrum. \label{figure_spectrum}}
\end{figure}

To confidently add spectral lines it is important to verify the
wavelength calibration and to correct for any systematic offsets
between the observed line centroids and the reference wavelengths from
the APED database. To fit the emission lines we first determined the
contribution of the continuum emission to the spectrum. A global fit
to line-free regions of HEG and MEG spectra using the Astrophysical
Plasma Emission Code model \citep[APEC, version 1.3,][]{smi2001}
allows a self-consistent treatment of the continuum for all lines.  We
defined the line-free regions as any part of the spectrum greater that
0.1~\AA\ in width that is devoid of spectral lines with emissivities
larger than $10^{-18}$ ph~cm$^3$~s$^{-1}$ (this is $\sim$1/1000 of
brightest line in the spectrum) in the APED database. We found a good
fit for a thermal continuum (consisting of bremsstrahlung, radiative
recombination, and two-photon emission) for a temperature of 16$\pm$1
keV (see Figure~\ref{figure_spectrum}), in good agreement with the
shock temperature of $15.4^{+5.3}_{-2.6}$ keV reported by
\citet{fuj1997} for an \citet{aiz1973} model fitted to the {\it
ASCA\/} spectrum of \exhya, and the maximum temperature of 20 keV
reported by \citet{muk2003} for a cooling flow model fitted to the
same {\it Chandra} HETG data. A two-temperature continuum model did
not significantly lower the $\chi^2$ of the fit. This is not
surprising since the highest temperatures dominate the emission
measure distribution in an accretion column cooling mainly by means of
thermal bremsstrahlung.

To determine if any offsets between the observed wavelengths and their
reference values exist we performed the following fit. We set the
distances between each consecutive pair of emission lines equal
to their separation based on the APED database and only fit the
normalization of each line and the overall wavelength offset. Each
line is modeled using a Gaussian with a FWHM equal to the
instrumental FWHM (0.012 \AA\ for HEG and 0.023 \AA\ for MEG).
This resulted in the following offsets for each of the orders
(predicted $-$ observed wavelength): $-0.0012(2)$, $-0.0020(2)$,
$-0.0008(2)$, and $0.0004(2)$ \AA\ for HEG$-1$, HEG$+1$, MEG$-1$, and
MEG$+1$, respectively. Although the offsets are small, they do
constitute $\sim$10\% of the HEG FWHM and $\sim$5\% of the MEG FWHM,
or about 30--50 \kms\ (at 10 \AA), and need to be corrected for.

Even more important than the absolute and relative wavelength
calibration is its stability during an observation; this appears to be
extremly stable ({\it Chandra} Proposers Observatory
Guide\footnote{http://cxc.harvard.edu/proposer/POG/index.html}).
Furthermore, the rotation and orbital periods of the EX Hya system are
not relevant timescales for the satellite (dither period is 707.1 sec
in the pitch direction and 1000 sec in the yaw direction).

\subsection{Binary}
To look for changing velocities associated with the binary period in
the {\it Chandra} HETG data for \exhya\ we first phase folded the
spectra \citep[using the ephemeris of][]{hel1992} and divided the
observation into phase intervals centered at $\phiorbit = 0.05, 0.10,
..., 0.95, 1.00$. with a width of $\Delta_\phiorbit = 0.25$, where
\phiorbit\ indicates the phase in the binary orbit. Note that the phase
intervals are overlapping.

Unfortunately, by splitting the observation into smaller phase
intervals none of the spectral lines have sufficient signal-to-noise
ratio to confidently measure changes in their wavelengths. We
therefore added all lines in each interval to accumulate enough counts
to accurately measure the central wavelength. We accomplished this by
the following procedure: 1) For each grating first-order arm we
determined the count-weighted average wavelength of the lines in
Table~\ref{table_lines}, $\bar{\lambda} = \sum_{i} c_i \lambda_{{\rm
ref,}i}/\sum_{i} c_i$, where $c_i$ is the number of counts in the
spectrum for a segment 2$\times$FWHM wide centered on $\lambda_{{\rm
ref,}i}$ (from Table~\ref{table_lines}) minus the number of counts
predicted by the model continuum (see \S3.1).  2) Each line in the
spectrum was shifted by an amount equal to its reference wavelength,
$\lambda_{{\rm ref},i}$ and by the wavelength calibration correction
(see \S3.1). 3) The wavelength-binned shifted spectrum was then
transformed to a velocity-binned spectrum according to $v_{{\rm
bin},i} = c (\lambda_{{\rm bin},i} - \lambda_{{\rm ref},i}) /
\lambda_{{\rm ref},i}$. 4) The shifted velocity-binned profile for
each spectral line was added to form one line profile. However, since
the bin size for each line was different we chose to rebin each
$v_{{\rm bin},i}$ to the bin size corresponding to $\bar{\lambda}$:
$\bar{v}_{\rm bin} = c (\lambda_{\rm bin} - \bar{\lambda}) /
\bar{\lambda}$. 5) Finally, we measured the central velocity for this
composite line profile (CLP) .

\begin{figure}
\epsscale{0.7}
\plotone{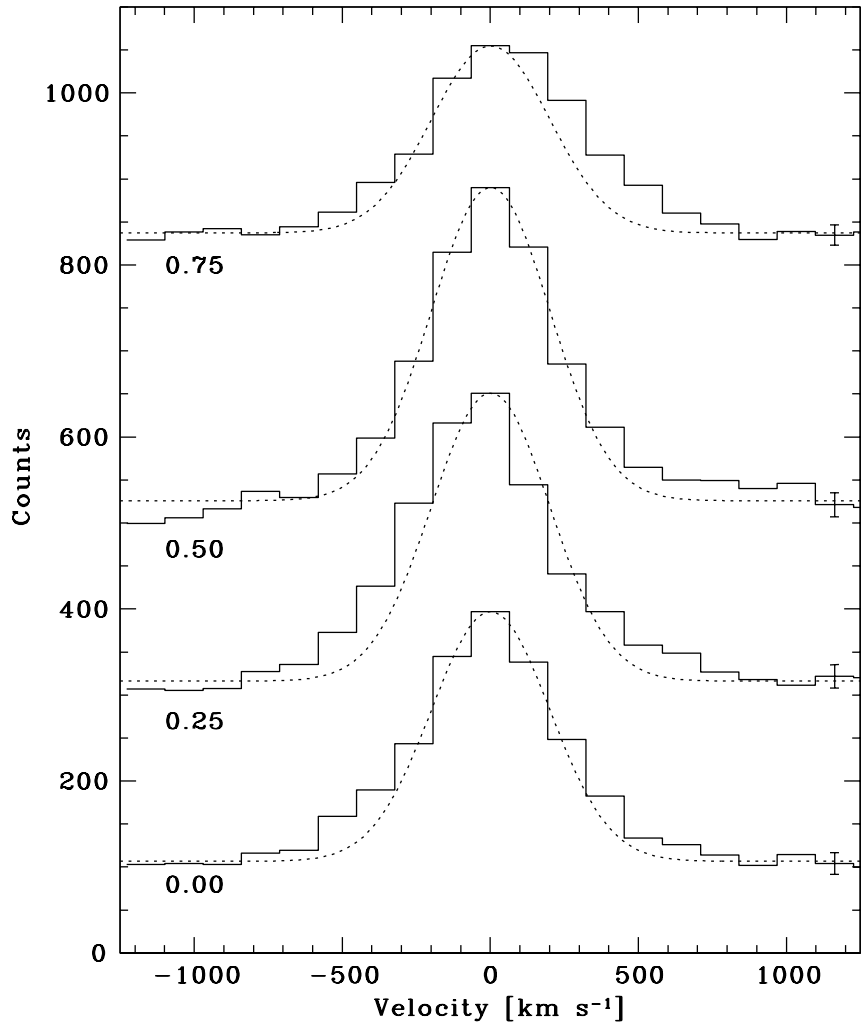}
\caption{Composite line profiles (CLPs) based on the first-order MEG
data for the non-blended lines of \exhya\ for binary phases 0.00,
0.25, 0.50, and 0.75 (shifted in the vertical direction by 200, 400,
and 720 counts, respectively, to separate the profiles). The average
error on the CLP bins is depicted in the last bin of each CLP.
Dotted line profiles represent the instrumental
profile of the CLP (normalized at the peak) at a constant velocity of
1.3~\kms. Number of counts in the 0.75 profile is
lower because this binary phase corresponds to the bulge absorption
feature in the \exhya\ light curve. \label{figure_profiles}}
\end{figure}

Figure~\ref{figure_profiles} shows the CLPs for binary phases
\phiorbit\ = 0.00, 0.25, 0.50, and 0.75. The change in CLP centroid is
clearly visible showing how powerful the combination of a large
effective area and high spectral resolution can be in the X-ray
regime.

\begin{figure}
\epsscale{0.7}
\plotone{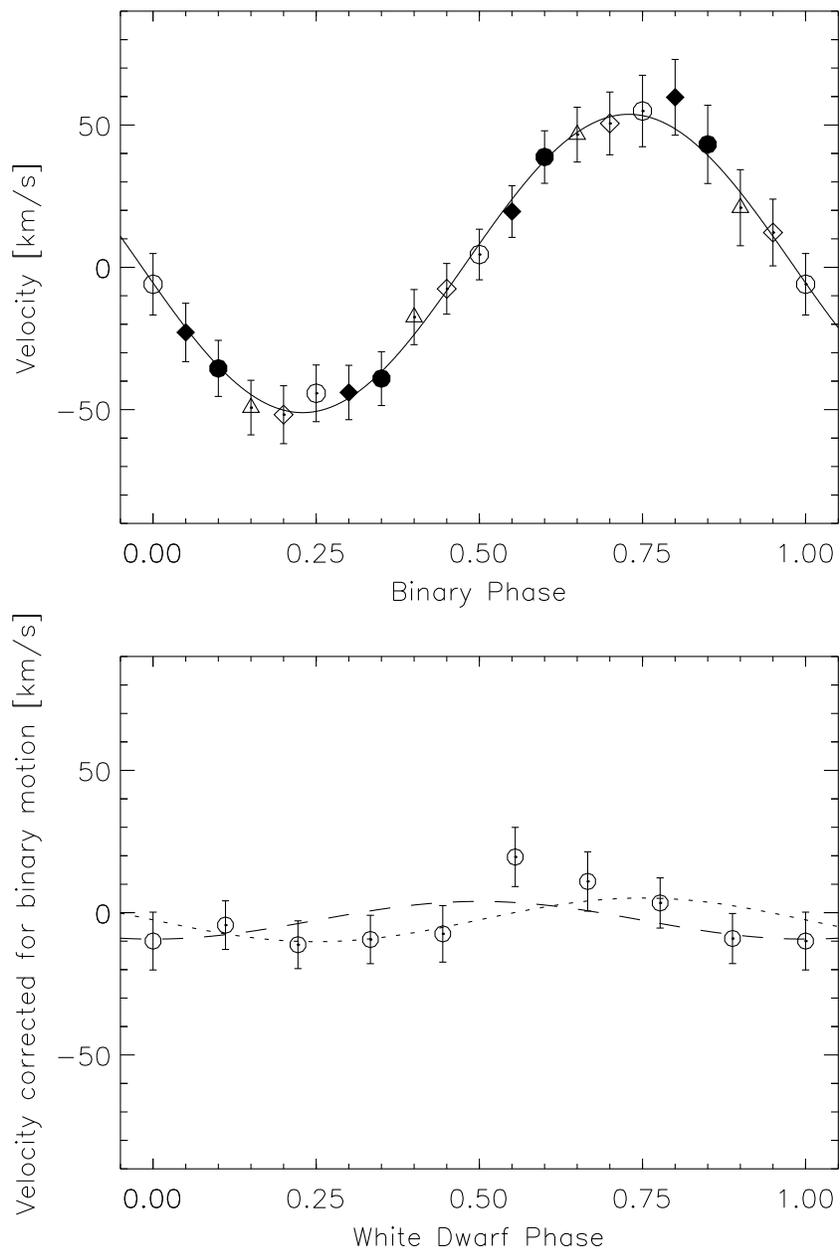}
\caption{Top panel: CLP radial velocity versus the binary period based
on the first-order MEG data. Solid curve represents the sinusoidal fit
to the data. The five sets of independent points have been indicated
by different symbols (see \S3.2). Bottom panel: CLP radial velocity
(corrected for the binary motion) versus the white dwarf spin
period. Dotted curve represents the best-fit velocity curve
corresponding to the rotation velocity of the white dwarf and the
dashed line represents the best-fit velocity curve corresponding to
the infall velocity in the accretion column (see \S3.3).
\label{figure_velocity}}
\end{figure}

The radial velocity shifts as a function of \phiorbit\ for the combined
orders MEG$-1$ plus MEG$+1$ are shown in the top panel of
Figure~\ref{figure_velocity}. We present the MEG data rather than the
HEG data because they have a factor of 2--10 times more counts per
spectral line, which improves our centroiding accuracy
even with the lower resolution of the MEG instrument (see also
\S5). Results from the HEG and the individual MEG orders are consistent
but with lower significance.  The figure clearly shows that there is a
systematic motion of the CLP with a period equal to that of the
binary.  The solid line represents the following fit to the data
points:
\begin{equation}
\label{eq_velocity}
v(\phi) = \frac{1}{\Delta}\int\limits_{\phi-\Delta/2}^{\phi+\Delta/2} \gamma + K
\sin \left ( 2\pi\frac{(\tilde{\phi}-\phi_0)}{\omega} \right ) {\rm d}\tilde{\phi},
\end{equation}
where $\gamma = 1.3\pm2.3~\kms$, $K = 58.2\pm3.7~\kms$ is the velocity
amplitude, $\phi_0 = 0.48\pm0.01$, $\omega$ is fixed to 1.0, and
$\Delta = 0.25$. Since the data points overlap, and are thus
correlated, we calculated $\chi^2$ for each of the five sets of
independent data points (indicated in
Figure~\ref{figure_velocity}). The $\chi^2$ per degree of freedom
(dof) for the independent data sets range from $\chi^2/({\rm dof =
1})$ = 0.1 to 1.0. It is important to note that the value of $\chi^2$
for a small number of dof can be different from the value one expects
to indicate a good fit in the case of a large number of dof, i.e.,
$\chi^2/{\rm dof} = 1.0$. The important statistic is $Q(\chi^2|\,{\rm
dof})$, which is the probability that the observed chi-squared will
exceed the value $\chi^2$. Acceptable values for $Q$, i.e., the
3$\sigma$ confidence interval, range from $\sim$0.0015 to
$\sim$0.9985. For the above values of $\chi^2$/dof, $Q$ ranges from
0.3 to 0.8.

We also fitted the data with 1) a constant velocity and 2) a sine
curve with the period not forced to the binary phase, i.e., $\omega$
not fixed to 1.0 in Eq.~\ref{eq_velocity}. The former fit yields
$\gamma = -2.8\pm2.3~\kms$, $\chi^2/({\rm dof = 3})$ ranges
from 13 to 19, and $Q \approx 7.0\times10^{-11}$, while the latter fit
yields a period of $\omega = 1.01\pm0.04$ times the binary period, and
values of $\gamma$, $K$, and $\phi_0$ are indistinguishable from the fit
described above. The latter fit has zero degrees of freedom and hence
a meaningless $\chi^2$.

The above velocity amplitude and phase offset, $K = 58.2\pm3.7~\kms$, and
$\phi_0 =0.48\pm0.01$ are in good agreement with those reported by
\citet[$K = 68\pm9~\kms$, $\phi_0 = 0.5$]{bre1980}, 
\citet*[$K = 90\pm28~\kms$, $\phi_0 = 0.55$]{cow1981}, 
\citet[$K = 58\pm9.3~\kms$, $\phi_0 = 0.52$]{gil1982}, 
\citet[$K = 69\pm9~\kms$, $\phi_0 = 0.53\pm0.03$]{hel1987}, 
\citet[$K = 85\pm9~\kms$, $\phi_0 = 0.54\pm0.02$, using FUV lines]{mau1999}, 
and \citet[$K = 59.6\pm2.6~\kms$, $\phi_0 = 0.48\pm0.05$, using UV lines]{bel2003}.

\subsection{White Dwarf Spin}
The bottom panel of Figure~\ref{figure_velocity} shows the CLP radial
velocity versus the white dwarf spin period using bins of size
one-third the spin period (i.e., $\Delta = 0.333$). The velocities
have been corrected for the average binary motion for each data point
based on the fit described in \S3.2.  To provide upper limits on the
rotation velocity of the white dwarf and the infall velocities in the
accretion column, we fitted sine curves to these data. We assumed that
the upper accretion column is pointing away from the observer at white
dwarf phase 0 (consistent with both the accretion curtain model
[\citealt{ros1988}] and the accretion pole occultation model
[\citealt{all1998}]) and that the cooler part of the lower accretion
column, where most of the spectral lines are formed, is not visible
throughout most of the spin cycle \citep{all1998}. The dotted curve in
the figure represents the best-fit rotation velocity of $9.2\pm5.0$
\kms\ ($\phi_0=0.50$) and the dashed curve represents the best-fit
infall velocity of $7.9\pm5.4$ \kms\ ($\phi_0=0.25$). The 3$\sigma$
upper limits on the rotation and infall velocities are thus $\sim$25~\kms .

\section{White Dwarf Mass}
Combining the orbital motion of the white dwarf obtained in \S3.2
with the orbital motion of the companion \citep[$K_2 =
360\pm35~\kms$]{put2003} and the eclipse duration  \citep{muk1998}
allows us to determine the white dwarf mass
$M_1$, companion mass $M_2$, and inclination $i$ of
the \exhya\ system.

Here we followed the approach taken by \citet{beu2003} using a
computer program kindly provided to us by K.\ Beuermann. Assuming that
the emission originates at or near the pole of the white dwarf, the
iterative method yields $M_{1} = 0.49\pm0.13$ ${\rm M}_\odot$, $M_2 =
0.078\pm0.014$ ${\rm M}_\odot$, $i = 77\fdg2\pm0\fdg6$. The white
dwarf mass $M_1$ corresponds to a white dwarf radius $R_1 =
1.0\pm0.2\times10^9$~cm \citep{woo1995}. If the emission originates
above the surface of the white dwarf the inclination of the system
decreases $\sim0\fdg1$ for every $0.1\times R_1$ in elevation above
the surface. These results agree within the errors with those found by
\citet{beu2003} and hence do not change any of the conclusions reached
in that paper.

Finally, we note that our estimate of the white dwarf mass, which is
based on the radial velocity of the emission lines in the
{\it Chandra\/} HETG spectrum, agrees with that of \citet{fuj1997} 
($M_1=0.48_{-0.06}^{+0.10}~{\rm M}_\odot$), which is based on the H-
to He-like emission line intensity ratios observed in the {\it ASCA\/}
spectrum. Our ``classical'' radial velocity result is applicable only
to high-inclination magnetic CV binaries, whereas theirs is applicable
to a wide range of polars and IPs, although it relies on theoretical
models of accretion column.

\section{Discussion}
Our radial velocity measurement in \exhya\ pushes the instrumental limit of
the {\it Chandra} grating instruments. The MEG instrument has an
effective area large enough to accumulate between 3000 and 4000 counts
in the CLP so that we can reliably find the centroid to an
accuracy of $\sim$0.5~m\AA, corresponding to an accuracy in
velocity space of 10--15 \kms\ (at $\sim$10~\AA). 
This is consistent with the statistical uncertainty for measuring
Gaussian line centroids derived by \citet*{lan1982}:
\begin{equation}
S_\mu = \sigma \left ( \frac{2}{\pi} \right )^{1/4} (\Delta x)_0^{1/2} \bar{\Delta},
\end{equation}
where $S_\mu$ is the statistical uncertainty in $\mu$ (the line
center), $\sigma$ is the instrumental resolution (FWHM/2.3548),
$(\Delta x)_0$ is the bin size expressed in terms of $\sigma$, and
$\bar{\Delta}$ is the average error expressed in terms of the peak
amplitude. For our MEG CLP this equation yields 0.5~m\AA, as
observed, where we have taken $\bar{\Delta} = 2/\sqrt{P}$, with $P$
the number of counts in the peak bin of the CLP.

Unfortunately the effective area of the HEG instrument is
significantly smaller (a factor of 3 at 10~\AA) and does not allow us
to improve over the MEG despite the better instrumental resolution ($S_\mu
= 0.5$~m\AA for HEG, similar to MEG). This means that for this \exhya\
data set we obtain similar velocity accuracies for HEG and MEG. To
fully exploit the HEG resolution, i.e., gain a factor of 2 in velocity
accuracy, a significantly longer observation would be required.

More velocity resolution is desirable to detect velocity changes
induced by the rotation of the white dwarf around it spin axis. Such
velocity signatures would provide information on the location and
velocity of the emitting plasma and on the location of the magnetic
pole. The lack of a pronounced velocity variation at the spin period
of the white dwarf in this HETG observation is expected. To see this,
note that the free-fall velocity onto the white dwarf $v_{\rm ff} =
(2GM_1/R_1)^{1/2}\approx 3600$ \kms, and the immediate post-shock
velocity $v=v_{\rm ff}/4\approx 900$ \kms. Also note that the majority
of the spectral lines in the HETG spectrum are formed at temperatures
of less than 2~keV, i.e., 0.1 times the post-shock temperature
$T=3GM_1m_{\rm H}\mu/8kR_1 \approx 20$ keV ($\mu = 0.615$ is the
mean molecular weight of a completely ionized plasma of solar
composition). From Figures~1 and 2 of \citet{aiz1973} we estimate the
velocity of the 2~keV gas to be $0.1\, v \approx 90$ \kms.  Assuming
that the accretion columns are perpendicular to the orbital plane, the
projected post-shock velocity $v\, \cos i \approx 18$
\kms. Furthermore, the maximum rotational velocity of the white dwarf
$v_{\rm rot} = 2 \pi R_1/ 67 {\rm min} \approx 15$ \kms. Both
velocities are smaller than the upper limits derived in \S3.3.

\section{Conclusions}
We have confirmed the orbital velocity of the white dwarf in the
cataclysmic variable \exhya\ using the {\it Chandra} HETG. This
measurement is the first detection of orbital motion in a CV binary
system based on X-ray spectroscopy. By adding various spectral lines
visible in the \exhya\ spectrum enough signal-to-noise ratio can be
accumulated to obtain velocities accurate to 10--15 \kms. No new
conclusions are drawn on the white dwarf mass (0.49$\pm$0.13 ${\rm
M}_\odot$).


\acknowledgments We thank Klaus Beuermann for his assistance in
determining the white dwarf mass.  We thank H.~Tananbaum for the
generous grant of Director's Discretionary Time that made possible the
{\it Chandra\/} observations of EX~Hya. Support for this work was
provided in part by NASA through {\it Chandra\/} Award Number
DD0-1004B issued by the {\it Chandra\/} X-Ray Observatory Center,
which is operated by the Smithsonian Astrophysical Observatory for and
on behalf of NASA under contract NAS8-39073. We acknowledge support
from NASA LTSA NAG5-3559 and Chandra grant GO2-3018x. NSB was
supported by NASA contract NAS8-39073 to the Smithsonian Astrophysical
Observatory (SAO) for the Chandra X-Ray Center (CXC). CWM's
contribution to this work was performed under the auspices of the
US Department of Energy by University of California Lawrence
Livermore National Laboratory under contract W-7405-Eng-48.

\clearpage


\clearpage

%
%
%
\begin{deluxetable}{llll}
\tablewidth{0pt}
\tabletypesize{\scriptsize}
\tablecaption{Strong \exhydra\ HETG spectral lines}
\tablehead{\colhead{Ion}&\colhead{Upper level}&\colhead{Lower level}&\colhead{Wavelength\tablenotemark{a}}\\
\colhead{}&\colhead{}&\colhead{}&\colhead{[\AA]}}
\startdata
\ion{Fe}{26}                  & $2p~^2P_{1/2,3/2}$           & $1s~^2S_{1/2}$   & \phn1.780 \\
\ion{S}{16}                   & $2p~^2P_{1/2,3/2}$           & $1s~^2S_{1/2}$   & \phn4.729 \\ 
\ion{S}{15}                   & $1s2p~^1P_{1}$               & $1s^2~^1S_{0}$   & \phn5.039 \\ 
\ion{Si}{14}                  & $2p~^2P_{1/2,3/2}$           & $1s~^2S_{1/2}$   & \phn6.182 \\ 
\ion{Si}{13}                  & $1s2p~^1P_{1}$               & $1s^2~^1S_{0}$   & \phn6.648 \\ 
\ion{Si}{13}\tablenotemark{b} & $1s2p~^3P_{1,2}$             & $1s^2~^1S_{0}$   & \phn6.687 \\ 
\ion{Fe}{24}                  & $4p~^2P_{1/2,3/2}$           & $2s~^2S_{1/2}$   & \phn7.989 \\ 
\ion{Mg}{12}                  & $2p~^2P_{1/2,3/2}$           & $1s~^2S_{1/2}$   & \phn8.421 \\ 
\ion{Mg}{11}                  & $1s2p~^1P_{1}$               & $1s^2~^1S_{0}$   & \phn9.169 \\ 
\ion{Mg}{11}\tablenotemark{b} & $1s2p~^3P_{1,2}$             & $1s^2~^1S_{0}$   & \phn9.231 \\ 
\ion{Ne}{10}                  & $3p~^2P_{1/2,3/2}$           & $1s~^2S_{1/2}$   &    10.239 \\ 
\ion{Fe}{24}                  & $3d~^2D_{5/2}$               & $2s~^2S_{1/2}$   &    11.176 \\ 
\ion{Fe}{24}\tablenotemark{c} & $3s~^2S_{1/2}$               & $2p~^2P_{3/2}$   &    11.432 \\ 
\ion{Fe}{23}                  & $2s3d~^1D_{2}$               & $2s2p~^1P_{1}$   &    11.736 \\ 
\ion{Ne}{10}\tablenotemark{d} & $2p~^2P_{1/2,3/2}$           & $1s~^2S_{1/2}$   &    12.132 \\ 
\ion{Ne}{9}                   & $1s2p~^1P_{1}$               & $1s^2~^1S_{0}$   &    13.447 \\ 
\ion{Fe}{18}                  & $2p_{1/2}2p_{3/2}^33d_{5/2}$ & $2p^5~^2P_{3/2}$ &    14.208 \\ 
                              & $2p^4(^1D)3d~^2D_{5/2}$      & $2p^5~^2P_{3/2}$ &           \\
\ion{Fe}{17}                  & $2p^5(^2P)3d~^1P_{1}$        & $2p^6~^1S_{0}$   &    15.014 \\ 
\ion{Fe}{17}                  & $2p^5(^2P)3d~^3D_{1}$        & $2p^6~^1S_{0}$   &    15.261 \\ 
\ion{Fe}{18}\tablenotemark{e} & $2p^4(^3P)3s~^2P_{3/2}$      & $2p^5~^2P_{3/2}$ &    16.005 \\ 
\ion{Fe}{17}                  & $2p^5(^2P)3s~^1P_{1}$        & $2p^6~^1S_{0}$   &    16.780 \\ 
\ion{Fe}{17}                  & $2p^5(^2P)3s~^3P_{1}$        & $2p^6~^1S_{0}$   &    17.051 \\ 
\ion{O}{8}                    & $2p~^2P_{1/2,3/2}$ 	     & $1s~^2S_{1/2}$   &    18.968 \\ 
\enddata   

\tablenotetext{a}{All wavelength errors are better than 0.004~\AA.}

\tablenotetext{b}{The unresolved He-like forbidden and
intercombination lines for Si and Mg are not considered in the
velocity analysis.}

\tablenotetext{c}{Blend consists of the \ion{Fe}{24}, \ion{Fe}{18}
$2p^4(^3P)4d~^2F_{5/2}$ to $2p^5~^2P_{3/2}$, and \ion{Fe}{22}
$2s2p_{1/2}3p_{3/2}$ to $2s^22p~^2P_{1/2}$ lines. We adopted
the wavelength for \ion{Fe}{24} since most flux seems to
be contained in this line.}

\tablenotetext{d}{Blend consists of the \ion{Ne}{10} Ly~$\alpha$
doublet, \ion{Fe}{17} $2p^5(^2P)4d~^1P_{1}$ to
$2p^6~^1S_{0}$, and \ion{Fe}{22} $2p(^3P)3d~^4D_{5/2}$ to
$2p^2~^2D_{5/2}$ lines.}

\tablenotetext{e}{Blend consists of the \ion{Fe}{18} and \ion{O}{8}
$3p~^2P_{1/2,3/2}$ to $1s~^2S_{1/2}$ lines.}

\label{table_lines}
\end{deluxetable}



\begin{thebibliography}{}
%
\bibitem[Aizu(1973)]{aiz1973} Aizu, K.\ 1973, Prg.\ Theor.\ Phys., 49, 1184
%
\bibitem[Allan et al.(1998)Allan, Hellier, \& Beardmore]{all1998}
Allan, A., Hellier, C.~H., \& Beardmore, A.\ 1998, MNRAS, 295, 167
%
\bibitem[Belle et al.(2003)]{bel2003} Belle, K.~E., Howell, S.~B., 
Sion, E.~M., Long, K.~S., \& Szkody, P.\ 2003, ApJ, 587, 373
%
\bibitem[Beuermann et al.(2003)]{beu2003} Beuermann, K., Harrison,
Th.~E., McArthur, B.~E., Benedict, G.~F., \& G\"ansicke, B.~T.\ 2003,
A\&A, 412, 821
%
\bibitem[Breysacher \& Vogt(1980)]{bre1980} Breysacher, J., \& Vogt,
N.\ 1980, A\&A, 87, 349
%
\bibitem[Brown et al.(1998)]{bro1998}Brown, G.~V., Beiersdorfer, P.,
Liedahl, D.~A., Widmann, K., \& Kahn, S.~M.\ 1998, ApJ, 502, 1015
%
\bibitem[Brown et al.(2002)]{bro2002}Brown, G.~V., Beiersdorfer, P.,
Liedahl, D.~A., Widmann, K., Kahn, S.~M., \& Clothiaux, E.~J.\ 2002,
ApJS, 140, 589
%
\bibitem[Cowley et al.(1981)Cowley, Hutchings, \& Crampton]{cow1981}
Cowley, A.~P., Hutchings, J.~B., \& Crampton, D.\ 1981, ApJ, 246, 489
%
\bibitem[Drake(1988)]{dra1988} Drake, G.~W.\ 1988, Can.\ J.\ Phys.\ 66, 586
%
\bibitem[Ericsson(1977)]{eri1977}Ericsson, G.~ W.\ 1977,
J.\ Phys.\ Chem.\ Ref.\ Data, 6, 3
%
\bibitem[Fujimoto \& Ishida(1997)]{fuj1997} Fujimoto, R., \& Ishida,
M.\ 1997, ApJ, 474, 774
%
\bibitem[Gilliland(1982)]{gil1982} Gilliland, R.~L.\ 1982, ApJ, 258, 576 
%
\bibitem[Hellier et al.(1987)]{hel1987} Hellier, C., Mason, K.~O.,
Rosen, S.~R., \& C\'ordova, F.~A.\ 1987, MNRAS, 228, 463
%
\bibitem[Hellier \& Sproats(1992)]{hel1992} Hellier, C., 
\& Sproats, L.~N.\ 1992, IBVS, 3724, 1
%
\bibitem[Kylafis \& Lamb(1982)]{kyl1982} Kylafis, N.~D., \& Lamb, D.~Q.\ 1982, \apjs, 48, 239
%
\bibitem[Landman et al.(1982)Landman, Roussel-Dupr\'e, \&
Tanigawa]{lan1982} Landman, D.~A., Roussel-Dupr\'e, R., \& Tanigawa,
G.\ 1982, ApJ, 261, 732 
%
\bibitem[Mauche(1999)]{mau1999} Mauche, C.~W.\ 1999, ApJ, 520, 822
%
\bibitem[Mauche(2002)]{mau2002} Mauche, C.~W.\ 2002, ASP 
Conf.~Ser.~261: The Physics of Cataclysmic Variables and Related Objects, 
eds.\ B.~T.~G\"ansicke, K.~Beuermann, and K.~Reinsch (San Francisco: ASP),
113
%
\bibitem[Mauche et al.(2001)Mauche, Liedahl, \& Fournier]{mau2001} Mauche,
C.~W., Liedahl, D.~A., \& Fournier, K.~B.\ 2001, ApJ, 560, 992
%
\bibitem[Mauche et al.(2003)Mauche, Liedahl, \& Fournier]{mau2003} Mauche,
C.~W., Liedahl, D.~A., \& Fournier, K.~B.\ 2003, ApJ, 588, L101
%
\bibitem[Mukai et al.(1998)]{muk1998} Mukai, K., Ishida, M., Osborne,
J., Rosen, S., \& Stavroyiannopoulos, D.\ 1998, ASP Conf.\ Ser.~ 137:
Wild Stars in the Old West, eds.\ S.~Howell, E.~Kuulkers, and
C.~Woodward (San Francisco: ASP), 554
%
\bibitem[Mukai et al.(2003)]{muk2003} Mukai, K., Kinkhabwala, A.,
Peterson, J.~R., Kahn, S.~M., \& Paerels, F.\ 2003, ApJ, 586, L77
%
\bibitem[Rosen et al.(1988)Rosen, Mason, \& C\'ordova]{ros1988}Rosen,
S.~R., Mason, H.~O., \& C\'ordova, F.~A.\ 1988, MNRAS, 231, 549
%
\bibitem[Smith et al.(2001)]{smi2001} Smith, R.~K., Brickhouse, N.~S., 
Liedahl, D.~A., \& Raymond, J.~C.\ 2001, ApJ, 556, L91
%
\bibitem[Sterken et al.(1983)]{ste1983} Sterken, C., Vogt, N., Freeth,
R., Kennedy, H.~D., Marino, B.~F., Page, A.~A., \& Walker, W.~S.~G.\ 1983,
A\&A, 118, 325
%
\bibitem[Vande Putte et al.(2003)]{put2003} Vande Putte, D., Smith,
R.~C., Hawkins, N.~A., \& Martin, J.~S.\ 2003, MNRAS, 342, 151
%
\bibitem[Vogt et al.(1980)Vogt, Krzeminski, \& Sterken]{vog1980} Vogt,
N., Krzeminski, W., \& Sterken, C.\ 1980, A\&A, 85, 106
%
\bibitem[Watson et al.(1978)Watson, Sherrington, \& Jameson]{wat1978}
Watson, M.~G., Sherrington, M.~R., \& Jameson, R.~F.\ 1978, MNRAS,
184, 79
%
\bibitem[Wood(1995)]{woo1995} Wood, M. A., 1995, in Lecture Notes on
Physics, eds.\ D.\ Koester, and K.\ Werner, LNP, 443, 41
%
%
%
%
%
\end{thebibliography}
\end{document}